\documentclass[aps,prl,twocolumn,showpacs,superscriptaddress]{revtex4}
\usepackage{amsmath}
\usepackage{amssymb}
\usepackage{color}
\usepackage{graphicx}
\usepackage{graphicx,amsmath}
\usepackage{bm}
\begin{document}

\title{Occurrence of flat bands in strongly correlated Fermi systems
and high-$T_c$ superconductivity of electron-doped compounds}
\author{V.~A. Khodel}
\affiliation{NRC Kurchatov Institute, Moscow, 123182, Russia}
\affiliation{McDonnell Center for the Space Sciences \& Department of Physics,
Washington University, St.~Louis, MO 63130, USA}
\author{J.~W. Clark}
\affiliation{McDonnell Center for the Space Sciences \&
Department of Physics, Washington University, St.~Louis, MO
63130, USA} \affiliation{Centro de Ci\^encias Matem\'aticas,
Universidade de Madeira, 9000-390 Funchal, Madeira,
Portugal}\author{K. G. Popov}\affiliation{Komi Science Center,
Ural Division, RAS, Syktyvkar, 167982,
Russia}\affiliation{Department of Physics, St.Petersburg State
University, Russia}\author{V. R. Shaginyan}
\affiliation{Petersburg Nuclear Physics Institute, NRC Kurchatov
Institute, Gatchina, 188300, Russia} \affiliation{Clark Atlanta
University, Atlanta, GA 30314, USA}

\begin{abstract}

We consider a class of strongly correlated Fermi systems that
exhibit an interaction-induced flat band pinned to the Fermi
surface, and generalize the Landau strategy to accommodate a
flat band and apply the more comprehensive theory to electron
systems of solids. The non-Fermi-liquid behavior that emerges is
compared with relevant experimental data on heavy-fermion metals
and electron-doped high-$T_c$ compounds. We elucidate how
heavy-fermion metals have extremely low superconducting
transition temperature $T_c$, its maximum reached in the
heavy-fermion metal CeCoIn$_5$ does not exceed 2.3 K, and
explain the enhancement of $T_c$ observed in high-$T_c$
superconductors. We show that the coefficient $A_1$ of the
$T$-linear resistivity scales with $T_c$, in agreement with the
experimental behavior uncovered in the electron-doped materials.
We have also constructed schematic temperature-doping phase
diagram of the copper oxide superconductor $\rm
La_{2-x}Ce_xCuO_4$ and explained the doping dependence of its
resistivity.
\end{abstract}
\pacs{71.10.Hf, 71.27.+a, 74.20.-z} \maketitle

An especially challenging task confronting present-day condensed matter
theory is explication of the elusive origin of the NFL behavior observed
in strongly correlated Fermi systems beyond a critical point where the
low-temperature density of states $N(T\to 0)$ diverges without breaking
any symmetry inherent of the ground state.  In homogeneous matter, the
case to be addressed in this article, such a divergence is associated
with the onset of a {\it topological} transition (TT) signaled by
the emergence of an additional root $p=p_b$ of the condition
\cite{lifshitz,volovik}
\begin{equation}
\epsilon(p,\lambda_t; T=0)=0
\label{top}
\end{equation}
for vanishing of the single-particle energy measured from  the chemical
potential $\mu$, otherwise satisfied only by $p=p_F$, the Fermi
momentum.  Thus, if $\lambda_t$ is the critical coupling constant for
onset of the posited TT, the curve $\epsilon(p,\lambda_t)$ must
touch the axis $p$ at the bifurcation momentum $p=p_b$.  Accordingly,
at $ p\to p_b$ the group velocity $v(p)=\partial \epsilon(p,\lambda_t)
/\partial p$ vanishes as $\sqrt{\epsilon(p)}$ to yield
\begin{equation}
N(T\to 0)\propto T^{-1/2} .
\label{dt}
\end{equation}
In an original scenario advanced by I. M. Lifshitz more than
fifty years ago \cite{lifshitz} and recently applied to describe
electron systems of heavy-fermion metals \cite{tai1,tai2}, some
TTs are assumed to occur in systems of {\it noninteracting}
electrons moving in the external field of the crystal lattice.
In such a case, the quasiparticle occupation numbers adhere to
the discrete Fermi-liquid (FL) values $n(p)=0,1$ beyond the
Lifshitz topological transition (LTT) point, while the Fermi
surface becomes multi-connected. Topological transitions may
occur in systems of {\it interacting} fermions as well
\cite{llano}. Within the FL approach, the feasibility of the
Lifshitz (``bubble'' or ``pocket'') scenario for the
rearrangement of the Landau state was considered \cite{zb,shagp}
within the framework of the Landau equation \cite{lan}
\begin{equation} {\partial \epsilon(p,T)\over
\partial {\bf p}}={{\bf p}\over M}+\int f({\bf p},{\bf p}')
{\partial n(p',T)\over
\partial {\bf p}'}d\upsilon' \label{lan} \end{equation}
for the single-particle spectrum $\epsilon(p)$.  Here $M$ is the
mass of a free fermion, $d\upsilon=2d^3p/(2\pi)^3$ is the
three-dimensional volume element, \begin{equation}
n(p,T)=(1+e^{\epsilon(p)/T})^{-1} \label{qmd} \end{equation} is
quasiparticle momentum distribution, and $f({\bf p},{\bf p}')$
is a phenomenological interaction function. Eq.~(\ref{lan}) was
derived by Landau \cite{lan} from the equality
\begin{equation}
\int {\bf p}\, n(p)d\upsilon= M\int {\partial \epsilon(p)\over
\partial {\bf p}}n(p) d\upsilon
\label{gal}
\end{equation}
between the momentum of the system, moving with velocity $\delta
{\bf v}$, its mass flow stemming from Galilean invariance.  We
note that Eq. \eqref{gal} is also valid when the model of
heavy-fermion liquid is applicable, as it is in our present case
\cite{shagrep,book}. Equation (\ref{lan}) is then obtained upon
retaining only the leading terms in $\delta n(p)$ on both sides
of Eq.~(\ref{gal}) while invoking the FL relation
\begin{equation}
\delta \epsilon({\bf p},n)=\sum_{{\bf p}'} f({\bf p}, {\bf p}')\delta n(p').
\label{flr}
\end{equation}
As will be seen, Eq. \eqref{gal} allows one to introduce and
explore a different, {\it interaction-induced} type of
rearrangement of the Landau state, often called fermion
condensation and described more vividly as a swelling of the
Fermi surface \cite{ks,noz,vol}. This phenomenon has its genesis
in a proliferation of the number of roots of Eq.~(\ref{top}) to
form a {\it continuum}. (For recent articles on this topic, see
Refs.~\cite{lee,shagrep,mig100,vol2011,an2012,kats2014,shashkin,
vol2014,book}).

The emergence of such a phase transition in homogeneous matter can be
elucidated within the original Landau approach to FL theory, in which
the ground-state energy $E$ is postulated to be a functional of the
quasiparticle momentum distribution $n(p)$.  At $T=0$, the onset of
fermion condensation in homogeneous matter is attributed to the
occurrence of a nontrivial solution $n_*(p)$ of the variational
condition \cite{ks}
\begin{equation}
{\delta E(n,\lambda)\over \delta n(p)}-\mu=0 , \quad p\in [p_l,p_u] ,
\label{var}
\end{equation}
in a finite momentum interval surrounding $p_b$, the chemical
potential $\mu$ being determined from the Landau postulate that
the quasiparticle and particle numbers {\it coincide}.  It is
Eq.~(\ref{var}) that describes a flat band pinned to the Fermi
surface, here also referred to as the fermion condensate (FC).
Outside the FC domain $[p_l,p_u]$ the usual FL occupancies still
apply: $n_*(p)=1$ at $p<p_l$ and $n_*(p)=0$ at $p>p_u$. However,
the occupation numbers inside the FC, evaluated through
Eq.~(\ref{var}), change {\it continuously} between 1 and 0 with
increasing $p$. The volume occupied in momentum space by
quasiparticles with nonzero probability is no longer just
$p_F^3/3\pi^2$ but is instead given by the relation
\cite{mig100}
\begin{equation}
\int\limits_0^{p_l} d\upsilon +\int\limits_{p_l}^{p_u} n_*(p) d\upsilon
 = {p^3_F\over 3\pi^2}.
\label{ll}
\end{equation}

We now turn to the demonstration that Landau equation
(\ref{lan}) is to be modified in dealing with Fermi systems
harboring a FC. It is revealed by examination of analytic
properties of solutions of this equation in systems where the
interaction function $f({\bf p},{\bf p}')$ has no singularities
in momentum space. For this case, it has been established
\cite{shagp} that the solutions of Eq.~(\ref{lan}) are in fact
{\it analytic} functions of momentum $p$ in the full momentum
space. However, this property is lost if the system hosts a FC.
Indeed, the left side of Eq.~(\ref{var}) is nothing but the
quasiparticle energy measured from the Fermi surface, which
vanishes identically in the FC domain. Evidently, the FC domain
cannot occupy  the full momentum space
 -- a fact confirmed in analytically soluble models
of fermion condensation. Thereby we arrive at the strong
conclusion that the single-particle spectrum $\epsilon(p,n_*)$
of the problem with FC present must be a {\it nonanalytic}
function of momentum $p$, for if an any analytic function
vanishes identically in some domain, it  must vanish everywhere.
This implies that in the case where the interaction function $f$
has no singularities in momentum space, the FC solutions
$\epsilon(p,n_*)$ {\it cannot meet} Eq.~(\ref{lan}), and
therefore in systems with a FC, Landau equation (\ref{lan}) must
be modified to allow for such an eventuality. This modification
can be made on the basis of Eq.~(\ref{gal}) along the same lines
as Eq.~(\ref{lan}) was obtained in Landau theory. The profound
difference is that the system harboring a FC is, in fact, a {\it
two-component} system, as made quite evident in Eq.~(\ref{ll}).
Correspondingly, variations of the quasiparticle momentum
distribution $n(p)$ lead to the closed equation
\begin{equation}
0={{\bf p}\over M}+\int f({\bf p},{\bf p}') {\partial n_*(p')\over
\partial {\bf p}'}d\upsilon',\quad p,p'\in [p_l,p_u]
\label{grfc}
\end{equation}
for determining $n(p)=n_*(p)$ inside the FC domain derived with accounting
for the fact that outside the FC region,  $dn_*(p)/dp=0$.   In arriving at
this relation, we have ensured consistency with the central feature of
the FC-inhabited state that the group velocity {\it vanishes identically}
in the interval $[p_l,p_u]$, whose range is found with the aid of the
requirement $n_*(p)<1$.  The quasiparticle spectrum of the {\it normal}
component obeys the Landau-type equation
\begin{equation}
{\partial \epsilon(p)\over \partial {\bf p}}={{\bf p}\over M}
+\int f({\bf p},{\bf p}') {\partial n_*(p')\over
\partial {\bf p}'}d\upsilon' ,\quad  p\notin [p_l,p_u] ,
\label{grb}
\end{equation}
and in effect $p'\in [p_l,p_u]$, derived in the same way as
Eq.~(\ref{lan}) from Eq.~(\ref{gal}). Equation (\ref{grfc}) can
also be derived by differentiation of the basic equation
(\ref{var}) with respect to momentum ${\bf p}$, then adopting
the relation (\ref{flr}) between interaction-induced variations
of relevant quantities, rewritten as $\delta
\epsilon(p)=\delta\epsilon_0 +(f\delta n(p))$, where $\delta
\epsilon_0$ is the variation of the spectrum of noninteracting
quasiparticles. In homogeneous matter, where $ \partial
\epsilon_0/\partial {\bf p}$ is simply ${\bf p}/M$, we are led
to Eq.~(\ref{grfc}). The special convenience of this route lies
in the opportunity to generalize Eq.~(\ref{grfc}) for analysis
of the FC phenomenon in electron systems of solids. In doing so,
one needs to replace ${\bf p}/M$ by the corresponding derivative
$ \partial \epsilon_0/\partial {\bf p}$, evaluated, say, within
the tight-binding model, or within a more advanced microscopic
description of the electron spectrum $\epsilon_0({\bf p})$. A
definite mathematical signature identifies those interacting
many-fermion systems whose single-particle spectrum exhibits a
flat portion.  This is the topological charge (TC) of the
corresponding ground state.  For a system containing a FC, the
TC must take a {\it half-odd-integral} value, whereas the TC of
any unorthodox state featuring one or more Lifshitz pockets is
always {\it integral} \cite{vol}.

It is instructive to compare changes that occur in the
fundamental equation (\ref{lan}) beyond the point of fermion
condensation with those that occur at a second-order phase
transition.  In the latter case, some symmetry inherent in the
Landau ground state is broken, and a corresponding order
parameter comes into play (notably, the gap $\Delta(p)$ in
superfluid FL's), dramatically rearranging Eq.~(\ref{lan}) and
both the key FL quantities, i.e., the momentum distribution
$n_{FL}(p)=\theta(p_F-p)$ and the single-particle spectrum
$\epsilon_{FL}(p)=p_F(p-p_F)/M^*$, where $M^*$ is the effective
mass. In the FC topological phase transition no such order
parameter enters; nevertheless Eq.~(\ref{lan}) alters, being
replaced by the set of {\it two equations} (\ref{grfc}) and
(\ref{grb}), in harmony with the two-component character of the
phase containing a FC. Although the equations written above to
describe a system containing a FC were derived for zero
temperature, they do not strictly apply there, because states
involving a FC have the distinctive property of carrying {\it
nonzero residual entropy}
\begin{equation} S_0=-\sum\left(  n_*(p)\ln n_*(p)+(1-n_*(p))\ln
(1-n_*(p))\right) . \label{s0} \end{equation} Such behavior
would contradict the Nernst theorem $ S(T=0)=0$, if it continued
to $T=0$.  Resolution of the incipient contradiction must lie in
release of the excess entropy $S_0$ via one or another phase
transition associated, especially, with pairing or magnetic
ordering and occurring at a respective transition temperature
$T_c$ or $T_N$ as $T\to 0$.  Even so, the above equations do
yield an approximate description, in the same sense as Landau FL
theory furnishes a viable approximate theory of the
low-temperature properties of liquid $^3$He {\it despite} the
fact that this system undergoes a normal-to-superfluid
transition as $T \to 0$, simply because of its extremely low
critical temperature $T_c$ for termination of superfluidity.  By
the same token, as long as the corresponding $T_c$ is low
enough, Eqs.~(\ref{grfc}) and (\ref{grb}) support an acceptable
zeroth approximation for a system of fermions having a FC.

To properly address the finite-temperature case, the energy
$E$ must be replaced by the free energy $F$.  So modified, the
variational condition (\ref{var}) takes the Landau-like form
(cf.\ Eq.~(\ref{qmd}))
\begin{equation}
\epsilon(p,n_*)=T \ln {1-n_*(p,T)\over  n_*(p,T)} .
\label{noz}
\end{equation}
Taking the solution $n_*(p)$ of Eq.~(\ref{var}) or (\ref{grfc}) as
a zeroth approximation for $n_*(p,T)$ applicable at low $T$, we
infer that at finite temperatures the dispersion of the spectrum
in the momentum interval $[p_l,p_u]$ becomes proportional to
$T$ \cite{noz}.  (Since the change of the NFL occupation numbers
$n_*(p)$ incurred by finite $T$ in this region remains quite small,
we shall continue to use the term FC for the totality of the
corresponding single-particle states.) The temperature-dependent
generalization of the above FC equations is
obtained by inserting the derivative
\begin{equation}
{\bf v}({\bf p})\equiv {\partial \epsilon(p,T)\over
\partial {\bf p}}=-T {\partial n_*(p,T)/\partial {\bf  p}
\over n_*(p,T)(1-n_*(p,T)) }
\label{nozv}
\end{equation}
into the left side of Eq.~(\ref{grfc}) to yield, in the FC
domain,
\begin{equation}
-T {\partial n_*(p,T)/\partial {\bf  p} \over n_*(p,T)(1-n_*(p,T)) }
={{\bf p}\over M}+\int f({\bf p},{\bf p}') {\partial n_*(p',T)\over
\partial {\bf p}'}d\upsilon'.
\label{grfct} \end{equation} This equation holds in the FC
interval $ p\in [p_l,p_u]$, whose length shrinks with $T$,
vanishing at an upper critical temperature $T_f$. The presence
of the FC phase on the disordered side of the transition, where
standard FL theory should ordinarily exert dominion, results in
the breakdown of virtually all thermodynamic and kinetic
predictions of Landau FL theory.  Further discussion will be
focused on the magnitude and temperature dependence of the
low-$T$ resistivity in the disordered, FC phase, which may
represent the most conspicuous departures from FL theory. In
{\it elastic} scattering processes involving normal
quasiparticles and the FC, the condensate behaves as a system of
{\it impurities}, giving rise to a residual resistivity $\rho_0$
in {\it clean} metals, whose value {\it depends} on the pressure
$P$.  Experimental documentation of the consequent impact on
$\rho_0$ (an effect inconceivable within the textbook
understanding of kinetic phenomena in Fermi liquids) requires
high-quality samples without doping $x$ that introduces
substantial disorder.  In spite of this stricture, a profound
effect has been observed in measurements of the resistivity of
the heavy-fermion compounds CeCoIn$_5$ and CeAgSb$_2$
\cite{sidorov,onuki}. The latter compound undergoes an
antiferromagnetic transition, whose critical temperature
declines under pressure.  On the ordered side of this
transition, $\rho(T)$ demonstrates the usual FL behavior
$\rho(T)=\rho_0+A_2T^2$, with residual resistivity $\rho_0$
below 0.5 $\mu\Omega\, {\rm cm}$.  However, with increasing $P$
this familiar behavior of $\rho(T)$ is disrupted.  As $P$
approaches the value $P_c$ at which magnetic ordering is
destroyed, the residual resistivity $\rho_0$ abruptly soars
upward, reaching values of about 50~$\mu\Omega\,{\rm cm}$ on the
disordered side of the transition \cite{onuki}.  We submit that
such a jump of $\rho_0$ amounts to irrefutable evidence for the
presence of a FC on the disordered side of the phase transition.

The heavy-fermion metal CeCoIn$_5$ proves to be a superconductor
on the ordered side of the corresponding phase transition.  We
contend that in such systems, NFL behavior of $\rho(T)$ on the
disordered side of this transition is primarily associated with
{\it inelastic} scattering processes in which FC quasiparticles
are generated or eliminated {\it singly} so as to augment the FL
formula for $\rho(T)$ with a term linear in $T$, whose magnitude
$A_1$ is proportional to the FC range $L_f$
\cite{kss,kz1,cecoin5}: \begin{equation} A_1(x)\propto L_f(x).
\label{a1} \end{equation} The low-$T$ resistivity $\rho(T,P=0)$
found experimentally for the normal state of CeCoIn$_5$ is in
fact consistent with this behavior \cite{paglione}.  When the
pressure $P$ is raised to a critical value $P^*\simeq 1.6~{\rm
GPa}$, it is found that in a narrow pressure diapason around
$P^*$, CeCoIn$_5$ exhibits a crossover to Landau-like behavior
$\rho(T)=\rho_0+A_2T^2$. Concomitantly, the residual resistivity
$\rho_0$ is reduced tenfold in a topological transition from a
state with a FC to one without, dropping to a very small value
around $0.2~\mu\Omega\,{\rm cm}$ that is due solely to impurity
scattering. The crucial difference between the effects of the FC
and the impurities that are present lies in the fact that the
former -- belonging as it does to the system of electrons --
contributes to the resistivity {\it entirely by virtue of
Umklapp processes} occurring in the crystal lattice, which
spoils electron momentum conservation.  However, these processes
play no significant role in the thermal conductivity. One is led
to the conclusion that the Lorenz number
\begin{equation}
L_0=\lim _{T\to 0} \kappa(T)/( T\sigma(T))  ,
\end{equation}
evaluated at low pressures $P<P^*$ where the FC is present, must
be suppressed compared with the textbook value $L_0=\pi^2/3e^2$.
It is just such a violation of the Wiedemann-Franz law that
notoriously occurs in CeCoIn$_5$  \cite{paglione}.

\begin{figure}[!ht]
\begin{center}
\includegraphics [width=0.47\textwidth]{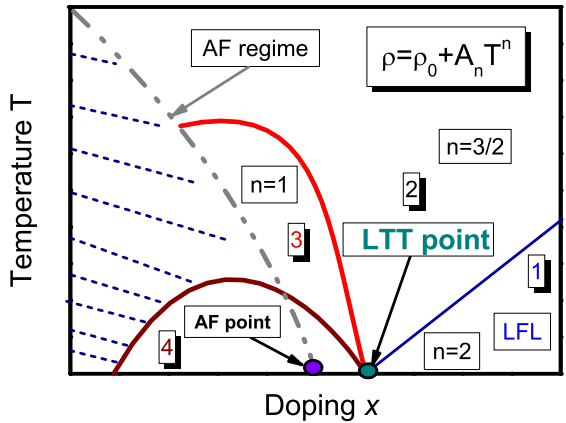}
\end{center}
\caption{Schematic temperature-doping $T-x$ phase diagram of
$\rm La_{2-x}Ce_xCuO_4$. The exponents $n$ of the resistivity
$\rho=\rho_0+A_nT^n$ are shown for three different regimes, that
appear around the LTT point, labeled 1,2,3. The Landau Fermi
liquid regime with $\rho(T)\propto A_2T^2$ dependence is labeled
by LFL. The transition areas between the corresponding regimes
are indicated by solid curves. Both the LTT point at $x=x_{LTT}$
and the antiferromagnetic point under the superconductivity dome
(labeled 4) are indicated by arrows. The onset of
antiferromagnetism is traced by the dash-dot-dot line.}
\label{fig1}
\end{figure}

These new insights into the observed NFL behavior of $\rho(T)$
may be elaborated as follows.  Replacement of a normal particle
in a scattering diagram by a FC quasiparticle entails
replacement, in the formula for the resistivity $\rho(T)$, of
the $T$-independent density of states $N_{FL}$ of Fermi-liquid
theory by the FC density of states $N(T)\propto T^{-1}$, thereby
promoting linear dependence of $\rho(T)$ on $T$. On the other
hand, at $T>T_f$ where the effect of the LTT point still
persists, the quantity $N_{FL}$ is replaced by $N(T)\propto
T^{-1/2}$ in accordance with Eq.~(\ref{dt}), yielding
$\rho(T)\propto T^{3/2}$. Such behavior closely resembles the
NFL behavior $\rho(T)\propto T^{1.5\pm 0.1}$ revealed in
measurements of the resistivity of the normal state of
CeCoIn$_5$ as well as in {\it electron-doped} high-$T_c$
superconductors \cite{armitage,greene1}.  Moreover, the
experimental $T-P$ phase diagram of CeCoIn$_5$ shown in Fig.~4
of Ref.~\cite{sidorov} and the experimental $T-x$ phase diagram
of $\rm La_{2-x}Ce_xCuO_4$ belonging to the LCCO family
presented in Fig.~1 of Ref.~\cite{greene1} are very much alike.
This empirical fact casts serious doubt on the premise that the
Kondo effect bears sole responsibility for the NFL behavior of
the resistivity $\rho(T)$ of heavy-fermion metals, since the
Kondo effect is absent in electron-doped compounds. In both of
these compounds, release of the residual entropy $S_0$ occurs
through the antiferromagnetic phase transition that can be
replaced by the superconducting phase transition at lowering
temperatures, as it is seen from Fig. \ref{fig1}.

Let us now discuss the enhancement of the superconducting
critical temperature $T_c$, attributed to critical
spin-fluctuations in Refs. \cite{taillefer,greene1}. Within  the
flat-band scenario,  this enhancement is  estimated on the base
of the standard BCS equation
\begin{equation}
1 =  {\cal V} \int\tanh \left(\epsilon(p,T_c)/
2T_c\right)/ 2\epsilon(p,T_c) d\upsilon.
\label{bcse}
\end{equation}
(For simplicity, we ignore the momentum dependence of the
pairing interaction, replacing it by the constant ${\cal V}$).
Upon manipulations based on  Eq. (\ref{noz}) the BCS Eq.
(\ref{bcse}) takes the form
\begin{equation}
1=-0.5{\cal V}[\alpha\eta n/T_c+N_n(0)\ln\left(\Omega_D/ T_c\right)]
\label{tc}
\end{equation}
with  total electron density  $n$, the dimensionless FC
parameter $\eta=(p_u-p_l)/p_F$ and numerical factor
$\alpha=O(1)$. In obtaining this result we retain  merely
leading terms  that diverge at $T_c\to 0$. The first  term in
square brackets  comes from momentum integration over the FC
region. The second,  usual BCS one is associated with
contributions from bands where normal quasiparticles reside. It
contains the familiar density of states $N_n(0)\propto
p_FM^*_n/\pi^2$. In the first approximation, the BCS term can be
neglected, and  then  from Eq. (\ref{tc}) one finds
\begin{equation}
T_c(x)\propto\eta(x), \label{tcx}
\end{equation}
i.e.  critical temperature $T_c$ turns out to be  linear
function of the FC parameter $\eta$. Comparing Eqs. (\ref{tcx})
and (\ref{a1}), we infer that both the $A_1$ term in the
resistivity $\rho(T)$ and critical temperature $T_c$ change with
input parameters $P$ and $\eta$  proportionally to the FC
parameter $\eta$.  Thus, the theoretical ratio $T_c/A_1$ is
approximately independent of the input, in agreement with the
experimental behavior uncovered in the electron-doped materials
LCCO and PCCO \cite{greene1,taillefer}.

We note in passing that the departure from the bare mass $M$ of
the effective mass $M^*_n$, extracted from experimental data on
the specific heat,  might be significant, as e.g. in the
heavy-fermion metals. This deviation  plays important role in
the magnitude of $T_c$. To demonstrate let us slightly
facilitate Eq. (\ref{tc}) by excluding the effective  coupling
constant ${\cal V}$ with the replacement of  the unity on the
l.h.s. of Eq. (\ref{tc}) from  BCS relation $1=-0.5{\cal
V}N_n(0)\ln\left(\Omega_D/T_c^{\rm BCS}\right)$ that yields
\begin{equation}
{T_c\over\epsilon^0_F}\ln \left(T_c/T_c^{BCS}\right)=\alpha \eta{M\over M^*_n}
\label{res1}
\end{equation}
with $\epsilon^0_F=p^2_F/2M$. The BCS case $T_c= T_c^{BCS}$ is
realized  provided the FC density $\eta\to 0$. Curiously, the
BCS situation also takes place in the heavy fermion metals, even
with the flat bands, because of the smallness of the ratio
$M/M^*_n\simeq 10^{-2}-10^{-3}$. This  explains how
heavy-fermion metals have extremely low $T_c$, its maximum
reached in the heavy-fermion metal CeCoIn$_5$ does not exceed
2.3\, K. In dealing with other electron systems where the ratio
$M/M^*_n$ is not  as small as in the heavy-fermion metals the
situation changes. Assuming  the needed attraction to come  from
the electron-phonon exchange  and  choosing  $M/M^*_n=1$,
$\alpha=0.1$,  $\eta=0.1$, one finds
\begin{equation}
T_c\simeq 0.3*10^{-2}\epsilon^0_F .
\label{tcfl}
\end{equation}
that explains the enhancement of $T_c$ observed in high-$T_c$
compounds and, particularly, in electron-doped high-$T_c$
materials. From Eq. (\ref{tcfl}) we infer that  the
superconducting phase transition in 2D electron gas of MOSFETs
and  some heterostructures  where the needed attraction in the
Cooper channel is furnished by the electron-phonon exchange may
have sufficiently high $T_c$ provided  the FC onset occurs near
a critical density, at which the  electron effective mass
diverges \cite{shashrev}. Noteworthy, in the heavy-fermion metal
CeCoIn$_5$, the BCS and FC contributions to $T_c$ are rendered
comparable by virtue of the characteristic effective-mass
enhancement.  This entails, in turn, a marked extension of the
boundary between SC and Landau Fermi liquid (LFL) regimes in the
$T-P$ phase diagram.

\begin{figure}[!ht]
\begin{center}
\includegraphics [width=0.47\textwidth]{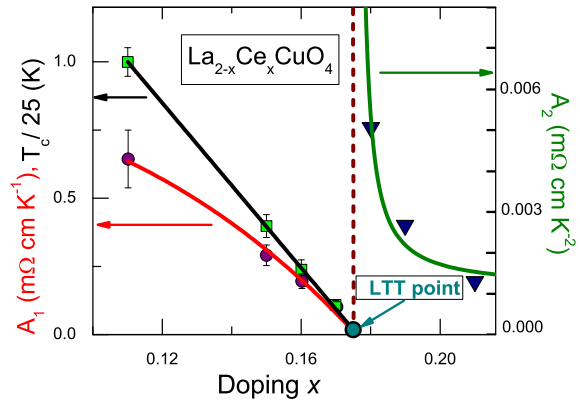}
\end{center}
\caption{Doping dependence of $T_c(x)$, $A_1(x)$ and $A_2(x)$ in
zero field \cite{greene1}. The left vertical axis shows the
coefficient $A_1$ and $T_c$ divided by 25, while the right
vertical axis shows the coefficient $A_2$. In the LFL regime, as
the system approaches the LTT point, shown by the arrow, and the
doping $x\to x_{LTT}$ from higher doping, the coefficient $A_2$
increases in accordance with Eq. \eqref{LTT}, as it is depicted
by the solid curve.} \label{fig2}
\end{figure}

It is a remarkable feature of models of fermion condensation
\cite{ks,noz,physrep} that nontrivial solutions of
Eq.~(\ref{var}) carrying a residual entropy $S_0$ often emerge
just beyond the LTT point where the resistivity $\rho(T)$ varies
as $T^{3/2}$. As a result, near the LTT point the two NFL
regimes, proportional respectively to $T^{3/2}$ and $T$, are
situated adjacent to each other, as it is shown in the schematic
phase diagram \ref{fig1}. It is seen from Fig. \ref{fig1}, that
four different regimes for the resistivity behavior come into
play in the immediate vicinity of FC onset: (1) the LFL regime
$\rho(T)\propto T^2$, (2) the LTT-point regime $\rho(T)\propto
T^{3/2}$, (3) the FC regime $\rho(T)\propto T$, and (4) the
high-$T_c$ superconducting regime with $\rho=0$. To confirm our
analysis of the phase diagram shown in Fig. \ref{fig1}, we
consider the behavior of the resistivity in the regions "1" and
"3". In the region "3", as it is seen from Fig. \ref{fig2} and
follows from the above discussion, the coefficient $A_1\to 0$ as
the doping $x$ approaches the LTT point from above, while in
accordance with Eq. \ref{res1} $T_c\to0$. On the other hand, as
the doping $x$ tends to the LTT point from below, the
coefficient $A_2$ diverges, for in the LFL regime due to the
conservation of the Kadowaki–Woods ratio one has $A_2\propto
(M^*)^2$ \cite{kadw,khodshuck}. As a result, one has
\cite{shagrep,book}
\begin{equation}
(M^*(x))^2\propto A_2\simeq
\left(a_1+\frac{a_2}{x-x_{LTT}}\right)^2, \label{LTT}
\end{equation}
where $a_1$ and $a_2$ are constants and $x_{LTT}$ is the doping
$x$ at which the LTT point takes place. It seen from Fig.
\ref{fig2}, the right panel, that the behavior of $(M^*)^2$
given by Eq. \eqref{LTT} is in agreement with the experimental
facts.

In conclusion, prominent on the scene of this seminal subfield
of condensed matter physics are numerous versions of
Hertz-Millis-Moriya (HMM) theory that ascribe NFL behavior of
strongly correlated Fermi systems to quantum critical
fluctuations.  In stark contrast to these theories, experimental
studies of the class of such systems addressed herein furnish
clear evidence of the persistence of NFL behavior in regions far
enough from the lines $T_c$ and/or $T_N(x)$ that the influence
of critical fluctuations should be minimal.  The crux of the
matter is that the single-particle degrees of freedom are the
real playmakers for NFL behavior, and the traditional mission of
integrating them out is futile, as in throwing the baby out with
the bath water.

The authors are indebted to J. Paglione, E. Saperstein, and
M.~V.~Zverev for valuable discussions.  This work is supported
by NS-932.2014.2 and RFBR Grants: 13-02-00085, 14-02-00044.  VRS
is supported by the Russian Science Foundation, Grant
No.~14-22-00281. VAK thanks the McDonnell Center for the Space
Sciences for support.  JWC expresses his gratitude to Professor
Jos\'e Lu\'is da Silva and his colleagues at Centro de
Ci\^encias Matem\'aticas for generous hospitality and fruitful
discussions during summer residence at the University of
Madeira.

\end{document}